# The Galactic Center cloud G2 and its gas streamer

Oliver Pfuhl[1], Stefan Gillessen[1], Frank Eisenhauer[1], Reinhard Genzel[1,2,3], Philipp M. Plewa[1], Thomas Ott[1], Alessandro Ballone[1,4], Marc Schartmann[1,4], Andreas Burkert[1,4], Tobias K. Fritz[5], Re'em Sari[6], Elad Steinberg[6], Ann-Marie Madigan[3,7]

[1]Max Planck Institute for Extraterrestrial Physics, PO Box 1312, Giessenbachstr., 85741 Garching, Germany
[2]Department of Physics, Le Conte Hall, University of California, 94720 Berkeley, USA
[3]Department of Astronomy, B-20 Hearst Field Annex # 3411, University of California, 94720 Berkeley, USA
[4]Universitätssternwarte der Ludwig-Maximilians-Universität, Scheinerstr. 1, 81679 München, Germany
[5]Department of Astronomy, University of Virginia, Charlottesville, VA 22904-4325, USA
[6]Racah Institute of Physics, Hebrew University of Jerusalem, 91904 Israel
[7]Einstein Postdoctoral Fellow

*Abstract*
We present new, deep near-infrared SINFONI @ VLT integral field spectroscopy of the gas cloud G2 in the Galactic Center, from late August 2013, April 2014 and July 2014. G2 is visible in recombination line emission. The spatially resolved kinematic data track the ongoing tidal disruption. The cloud reached minimum distance to the MBH of 1950 Schwarzschild radii. As expected for an observation near pericenter passage, roughly half of the gas in 2014 is found at the redshifted, pre-pericenter side of the orbit, while the other half is at the post-pericenter, blueshifted side. We also present an orbital solution for the gas cloud G1, which was discovered a decade ago in L'-band images when it was spatially almost coincident with Sgr A*. The orientation of the G1 orbit in the three angles is almost identical to the one of G2, but it has a lower eccentricity and smaller semi-major axis. We show that the observed astrometric positions and radial velocities of G1 are compatible with the G2 orbit, assuming that (i) G1 was originally on the G2 orbit preceding G2 by 13 years and (ii) a simple drag force acted on it during pericenter passage. Taken together with the previously described tail of G2, which we detect in recombination line emission and thermal broadband emission, we propose that G2 may be a bright knot in a much more extensive gas streamer. This matches purely gaseous models for G2, such as a stellar wind clump or the tidal debris from a partial disruption of a star.

## *1. Introduction*

The gas cloud G2 was detected by Gillessen et al. (2012) in high-resolution images of the Galactic Center as an L'-band continuum source, with no apparent K-band counterpart, as well as a line emission source in near-IR hydrogen and helium recombination lines. G2 is on a highly eccentric orbit ($e \approx 0.98$) with a pericenter radius of roughly 20 light hours from the central massive black hole Sgr A*. Its nature as a dusty gas cloud (with or without a central star) was established in the discovery paper already by the observation of an increasing tidal shear developing over several years, which can be seen in position-velocity (pv) diagrams of the recombination line emission. Phifer et al. (2013) pointed out that the original orbit based on L'-band astrometry, with a pericenter passage mid 2013, would be biased due to the crowding and dusty background close to Sgr A*. Using the Brackett-γ emission from near-infrared integral field spectrographs yielded a slightly updated, and still more eccentric orbit with pericenter passage in April 2014 (Phifer et al. 2013, Gillessen et al. 2013b).

The detection of G2 attracted substantial interest, partly because its mass of a few Earth masses (Gillessen et al. 2012, Shcherbakov 2014) is comparable to the mass present in the accretion flow around Sgr A* (Yuan et al. 2003, Xu et al. 2006). If the gas motion is effectively circularized and a significant fraction of the gas accretes onto the black hole, G2 results in a significant accretion event, unfolding roughly over the next decade (Schartmann et al. 2012). So far, no increase in the emission of Sgr A* has been reported, although this is not necessarily expected yet. The loss of angular momentum and energy that is required to bring G2 from the pericenter distance of around 2000 Schwarzschild radii ($R_S$) down to a few 10 $R_S$ is governed by the viscous timescale, which is estimated to be years (Burkert et al. 2012, Schartmann et al. 2012, Moscibrodzka et al. 2012). Another prediction is the formation of a shock front since G2 moves with a speed of roughly Mach 2 through the hot gas of the accretion flow. Despite continuous monitoring no significant excess X-ray (Gillessen et al. 2012) or radio emission (Sadowski et al. 2013a, 2013b) from that process have been detected thus far (Haggard et al. 2014, Chandler & Sjouwerman 2014).

The origin of the gas in G2 is debated, and a number of models have been proposed (see Gillessen et al. 2014 for a discussion). The most fundamental difference between the models is, whether they place a compact hidden source (e.g. a faint star) at the center of G2 or not, which could provide a continuous source of gas supply to G2 on its orbit. Two models without a central source are
- G2 might be a clump formed from the winds of the massive stars in the clockwise disk of young stars (Paumard et al. 2006, Cuadra et al. 2006, Lu et al. 2009, Bartko et al. 2009, Schartmann et al. 2012, Burkert et al. 2012, Yelda et al. 2014). This model naturally explains why G2's orbit is coplanar with the disk. The stars S91 and IRS16SW are candidates, considering their orbital phases (Martins et al. 2006, Gillessen et al. 2009);
- G2 could be the debris of a star that underwent a partial tidal disruption. Guillochon et al. (2014) suggested that a giant underwent a grazing collision with Sgr A*, leading to a clumpy gas stream on a shorter period orbit than the original star.

Two models including a central source are
- G2 could be an evaporating protoplanetary disk around a young star (Murray-Clay & Loeb 2012). The production of free gas should increase due to the increasing tidal shear as G2 approaches Sgr A*, leading to a significant increase in Brackett-γ luminosity;
- G2 could be produced by a windy star, such as a T-Tauri star. The shock front between wind and ambient medium might explain the emission from G2 (Scoville & Burkert 2013). Also in these models, the luminosity of G2 should increase during the approach (Ballone et al. 2013). The most up-to-date simulation of that scenario is presented in De Colle et al. (2014).

Other models include a nova origin (Meyer & Meyer-Hofmeister 2012) with a ring-like ejection of material. Such a ring has been shown in Schartmann et al. (2012) to be able to reproduce the head-tail structure of G2. The model would need a fine-tuning of the eruption time, though. Miralda-Escudé (2012) proposed that a main sequence star collided with a stellar black hole, leading to a formation of a circumstellar disk, which in turn would resemble the model of Murray-Clay & Loeb (2012).

There are other, similar sources in the central arcsecond (Sitarski et al. 2014). Already a decade ago, Clénet et al. (2004a, 2004b, 2005) and Ghez et al. (2005) presented observations of another L'-band source close to Sgr A*, which we will call G1 in the following. At the time, it was visible almost on top of the position of Sgr A*, offset by around 100 mas to the Southwest. Also for G1, no K-band counterpart has been found, making it a similarly red source as G2. G1 is most likely a dusty, ionized gas cloud of moderate mass, very similar to G2.

Here, we present new observations of G2 from late August 2013, April 2014 and July 2014, which are discussed in section 2. Section 3 summarizes our observational results for G2. In section 4 we discuss our results on G2 and G1. Section 5 presents a speculative connection between the objects G2 and G1. Overall, we conclude in section 6 that G2 is part of a much larger gas stream. In the appendix we discuss the infrared excess of the star S2, present a test-particle simulation of G2 including a drag-force model and show the most recent spectrum of G2.

## *2. New Observations*

We have obtained three new epochs[1] of near-infrared integral field spectroscopy of the central arcsecond centered on Sgr A*, consisting of two moderately deep integrations in August/September 2013, and July 2014, plus a deeper integration obtained from end of March to beginning of May 2014. As was the case for our previous data sets, we used SINFONI (Eisenhauer et al. 2003, Bonnet et al. 2004) in its adaptive optics scale (12.5 × 25 mas/pix) and the H+K grating, yielding a spectral resolution of R ≈ 1500.
- The August 2013 data set was obtained during two visitor mode runs, with a total on-source integration time of 600 minutes, of which 380 minutes pass our quality cut, demanding that the FWHM of a star at 2.2μm in the reconstructed cube be less than 7 pix (87.5 mas). These data were obtained in natural guide star mode.
- The April 2014 data set was obtained during three visitor mode runs and some service mode time. We obtained a total on-source integration time of 1380 minutes, of which 1090 minutes pass the quality cut.  650 minutes were obtained in laser guide star mode, the rest in natural guide star mode.
- The July 2014 data was obtained in service mode with a total on-source integration time of 430min. These data were obtained in natural guide star mode.

The field of view of a single data cube is 0.8'' × 0.8''. We dithered in a quadratic pattern by half that size, such that the central region containing Sgr A* and G2 is covered in each cube and sampled at 12.5 mas/pix, yet a total area of roughly 1.2'' × 1.2'' is covered.

We applied our standard data reduction recipes for SINFONI, including sky subtraction, flat-fielding, bad pixel and distortion correction. The wavelength calibration uses emission line lamps, and is refined in a subsequent step, in which each object cube is shifted spectrally on a subpixel level to optimally match the atmospheric OH lines. We extract pv-diagrams with a slit width of 8 pixels (≈ 0.1'') along the G2 orbit based on Brackett-γ astrometry from Gillessen et al. (2013b). This approach of cutting the 3D-data cube is optimal for analyzing the orbital dynamics of G2, as the orbit is viewed more or less from behind, such that only one spatial coordinate is well resolved by our data, and changes along the line of sight velocity coordinate are pronounced. We use both the Brackett-γ emission and the Helium-I line emission for obtaining noise-weighted, coadded pv-diagrams. Unlike for the data from April 2013 (Gillessen et al. 2013b), we found that the Paschen-α emission is too much affected by the atmospheric features between the H- and K-band to be useful. We also re-extracted pv-diagrams for all other suitable epochs along the same orbit, since we previously used the orbit based on L'-band astrometry. In total, we can present ten pv-diagrams from 2004, 2006, 2008, 2010, 2011, 2012, 2013/04, 2013/08, 2014/04 and 2014/07.

---

[1] Based on observations collected at the European Southern Observatory, Paranal, Chile; programs  092.B-0088(AB), 092.B-0238(C), 092.B-0398(BD), 093.B-0217(F), 093.B-0218(AD).

## 3. Results

*3.1 The tidal disruption unfolds*

Our new data sets confirm the picture presented in Gillessen et al. (2012, 2013a, 2013b). We witness the ongoing tidal disruption of a gas cloud with an originally compact head (G2) and an extended tail (figure 1). The data from August 2013 show that compared to the April 2013 data (Gillessen et al. 2013b) the bulk of the emission has moved to even higher redshift and that the spread in velocity space has increased again. The mean redshift is around 2500 km/s, with a width of around 500 km/s. This data set is not as deep as the previous one, but we again detect hints of the emission of the gas at the post-pericenter side, i.e. on the blueshifted side.

In the data set from April 2014 the emission on the red side of the pv-diagram peaked at 2800 km/s with a width of FWHM= 640 km/s. On the blue side, we now measure the peak emission at -2300 km/s and a width of 850 km/s coincident (in location and slope in the pv-diagram) with the orbit at a high significance level (see Figure 1). This is gas that has passed pericenter. Its luminosity is now comparable to the one of the pre-pericenter, red-shifted side (see figure 1). This is exactly as expected for a tidally stretched gas cloud observed during pericenter passage.

The July 2014 data set is very similar to the one from the April 2014 data set, but is less deep and thus more affected by noise. Nevertheless, it shows clearly (and independently) from the previous data set, that we detect gas on the blue side of the orbit.

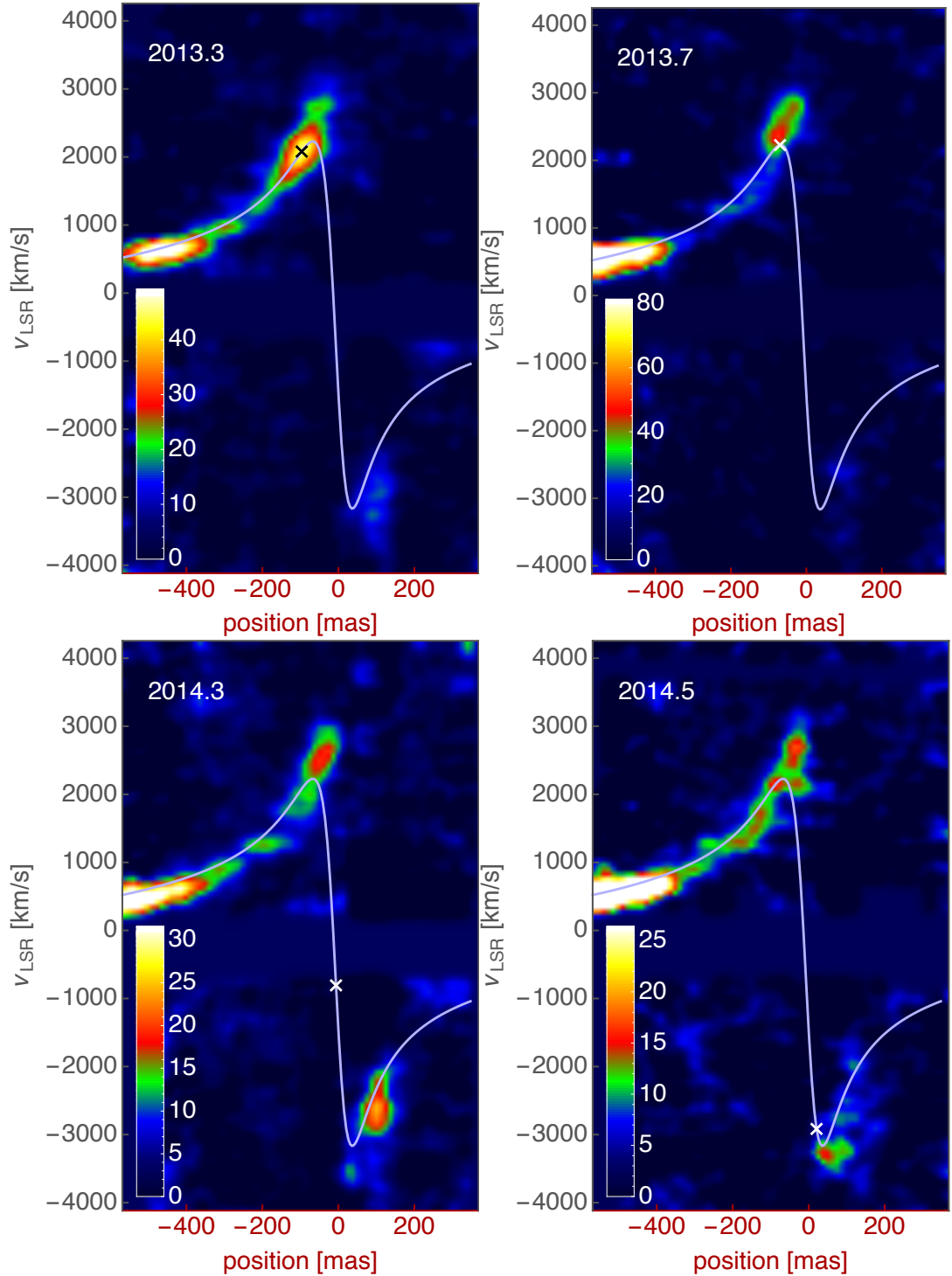

Figure 1: Comparison of the pv-diagrams from April 2013 (data already presented in Gillessen et al. 2013b), late August 2013, April 2014 and July 2014 (new data). The diagrams show the observed line-of-sight velocity as function of radial distance from SgrA*. The color scales state the observed luminosity in $10^{-6}$ $L_\odot$/pix$^2$, where a pixel covers a phase space region of 69 km/s × 12.5 mas). The scaling is adjusted in each map individually to optimally show the structure of the gaseous emission. The integrated luminosity is shown in figure 2 and discussed in section 3.2. The solid line corresponds to the Brackett-γ based orbit from Gillessen et al. (2013b), along which the pv-diagram is extracted. The cross marks the fitted position on the respective date. We have blended out the range between −660 km/s and +240 km/s to avoid emission from the mini-spiral (Paumard et al. 2004) visible at these wavelengths.

Given the complex and extended appearance (see figure 3), positions and radial velocities for the head of G2 would be ill defined in our new data sets. Hence, we continue to use the orbital parameters as given in Gillessen et al. (2013b). We note, however, that the pv diagrams in the 2013/08 and 2014 epochs would favor a (still) higher eccentricity than e = 0.976 (Gillessen et al. 2013b). Such a value was found by Phifer et al. (2013), who derived e = 0.981.

*3.2 The Brackett-γ luminosity of G2 has increased moderately*

We have measured the Brackett-γ flux of the head of G2 for both the April 2013 and the April 2014 epoch. Since the emission is spread out in the spatial as well as in the spectral domain, these measurements come with relatively large errors. The dominant error source is the definition of the source region, which is hard to define for a diffuse cloud. Hence, we obtain the errors by varying the extraction region.

We find that the head of G2 in 2013/04 had a Brackett-γ luminosity of $2.3 \pm 0.6 \times 10^{-3}$ $L_\odot$, when considering only the redshifted part of the emission. This value is consistent with all previous flux measurements (Gillessen et al. 2013b). The blueshifted part adds another $1.2 \pm 0.4 \times 10^{-3}$ $L_\odot$, yielding together thus a moderate but not yet statistically significant flux increase. In our 2014/04 data set we obtain $2.1 \pm 0.4 \times 10^{-3}$ $L_\odot$ for the redshifted side, and $2.4 \pm 0.6 \times 10^{-3}$ $L_\odot$ for the blueshifted side. The total brightness of G2 appears to have almost doubled between 2013/04 and 2014/04 thus (figure 2). However, this increase in flux is paralleled with a spread over more spatial and spectral channels, yielding a lower surface brightness and making observations increasingly challenging.

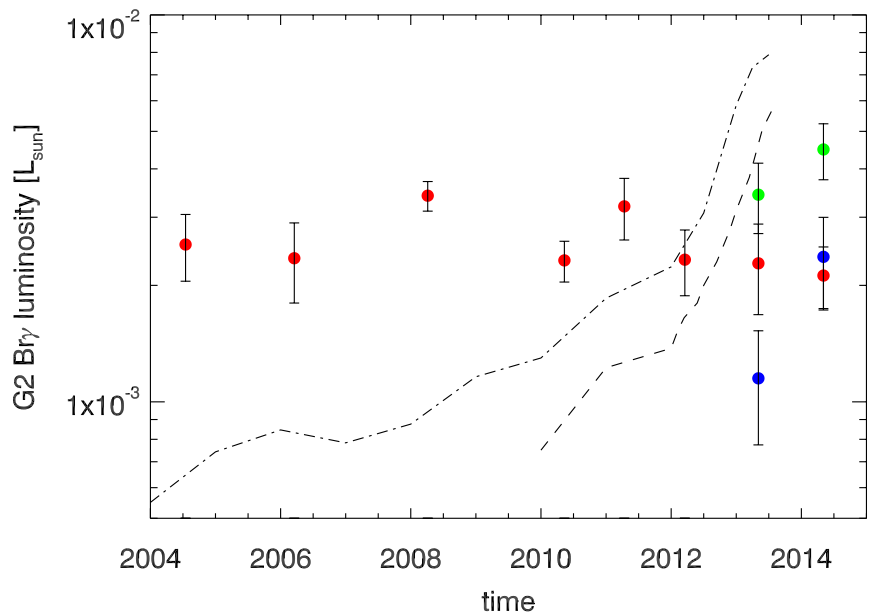

Figure 2: Evolution of G2's Brackett-γ luminosity. The red points mark the luminosity of the redshifted part of G2. For the 2013/04 and 2014/04 epochs, the blue points mark the luminosity of the blueshifted part. The green points show the sum of red- and blueshifted side. The dashed line shows the prediction from Murray-Clay & Loeb (2012), the dashed-dotted line the one from Ballone et al. (2013).

*3.3 The tail of G2*

In the 2013/04, 2014/04 and 2014/07 epochs we re-detect the tail emission that follows G2 along its orbit in the pv-diagram (figure 1). In the 2014 data sets it is actually difficult to separate G2 and the tail, because of the extreme tidal shear and because roughly half of the gas has moved to the blueshifted side.

We also extract the Brackett-γ map, which shows the tail in the spatial domain (figure 3). For this purpose we employ the same procedure as for the 2012 data set in Gillessen et al. (2013a), but scale the individual image planes such that tail emission has identical peak brightness with respect to the local background. Thus, the resulting maps are not photometrically correct any more, but they capture well in a single frame the structure of the tail, which extends over 25 spectral channels otherwise. These maps thus show that the gas emission of the tail is also unambiguously detected in our deep 2013 and 2014 integrations, and appears broadly consistent with the 2012 data.

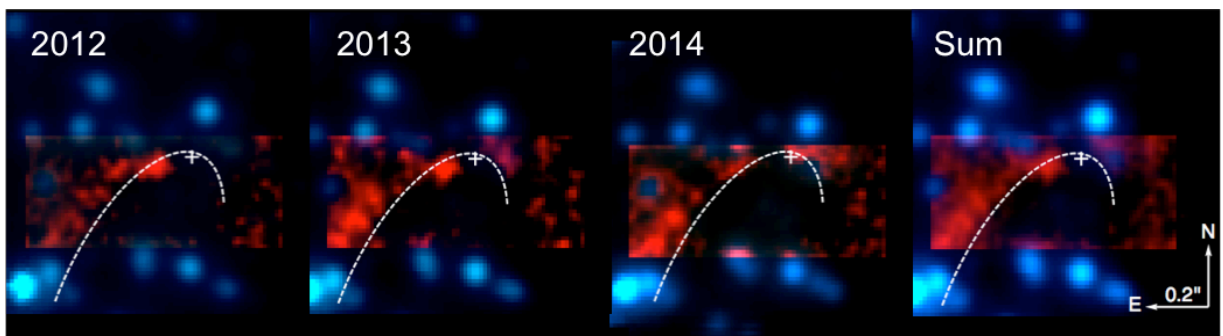

Figure 3: Evolution of the Brackett-γ emission of G2's tail in the image domain. The data set from 2012 was already presented in Gillessen et al. (2013a). The 2013 epoch is extracted from the 2013/04 data set, first presented in Gillessen et al. (2013b). The 2014 epoch is from the 2014/04 data set. The maps are generated as a sum of 25 channel maps each, scaled such that tail emission has identical peak brightness with respect to the local background. The right panel is the sum of the three others. The position of Sgr A* is marked with a cross, and the orbit of G2 is plotted as white dashed line.

The tail structure is also visible in our L'-band images (figure 4) obtained over the last decade with NACO (Lenzen et al. 1998, Rousset et al. 1998). The identification is more challenging due to the lower resolution at 3.8 μm and the high degree of confusion with either stars or other gaseous structures, but it is apparent in all high quality L'-band maps. Within the uncertainties, the L'-band luminosity of G2 remained constant at about $m_{L'}$ = 14.4 between 2003-2011. After 2011 G2 was too confused with other sources to obtain reliable photometry. The L'-band tail coincides with the Brackett-γ tail (figure 4, bottom right).

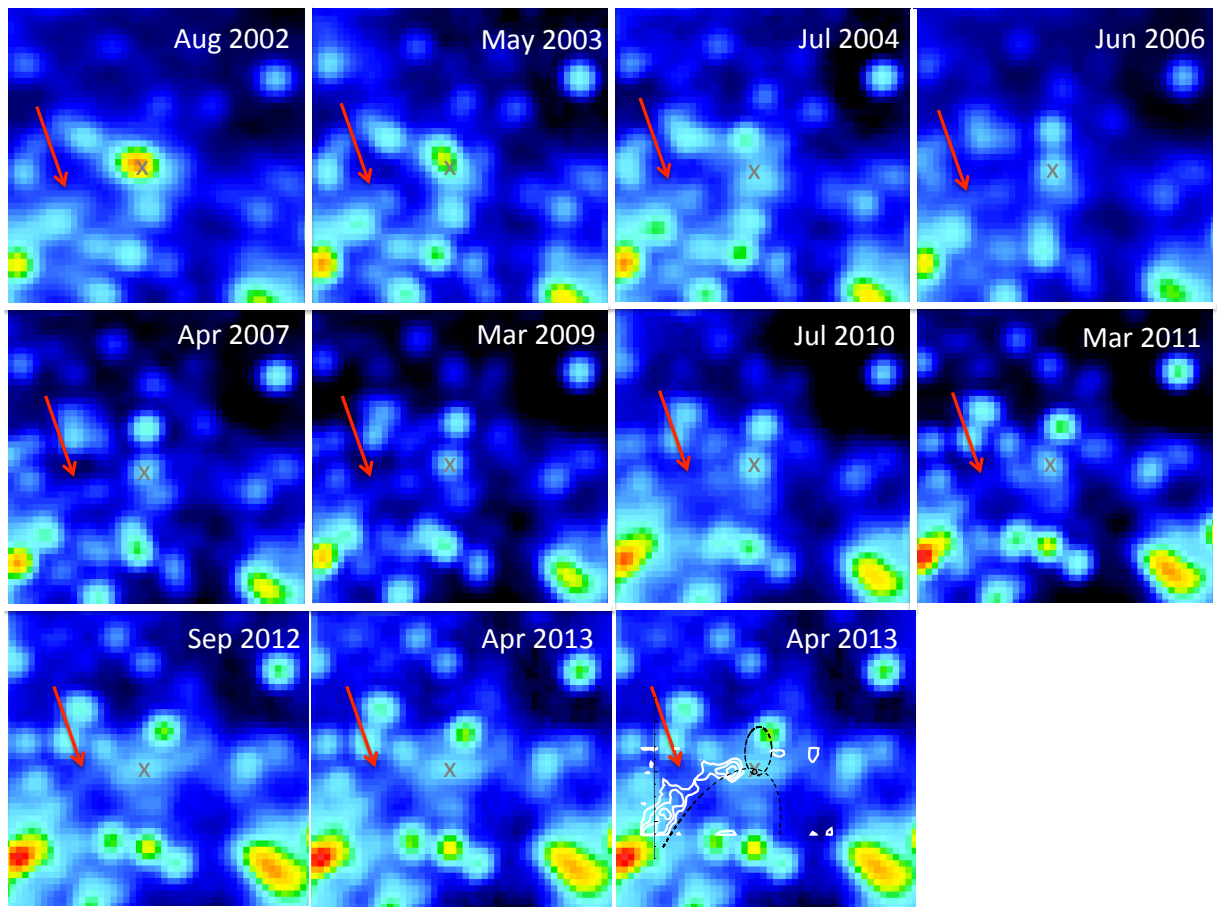

Figure 4: Time series of L'-band images of the central arcsecond. The arrow marks the tail structure following G2. The position of Sgr A* is marked with a cross. The rightmost panel in the bottom row shows on top of the 2013 image in contours the tail as presented in the right panel of figure 3, and for reference the orbits of S2 and G2.

## 3.4 The gas cloud G1

The pv-diagrams of the years 2004, 2006 and 2008 show some weak gas emission on the blueshifted, post-pericenter side, which apparently moves roughly along the predicted orbit of G2 (figure 5). Going back to the original SINFONI data cubes, it is apparent that this gas emission is spatially much more spread out than G2. It has a lower Brackett-γ line surface brightness, and it is difficult to define its borders both in the spatial dimensions and along the spectral axis. In figure 6 we show color composites of the line emission for the 2006 and 2008 data sets, showing that this blue emission is more diffuse than G2. This is the main systematic uncertainty for determining its astrometry and radial velocities.

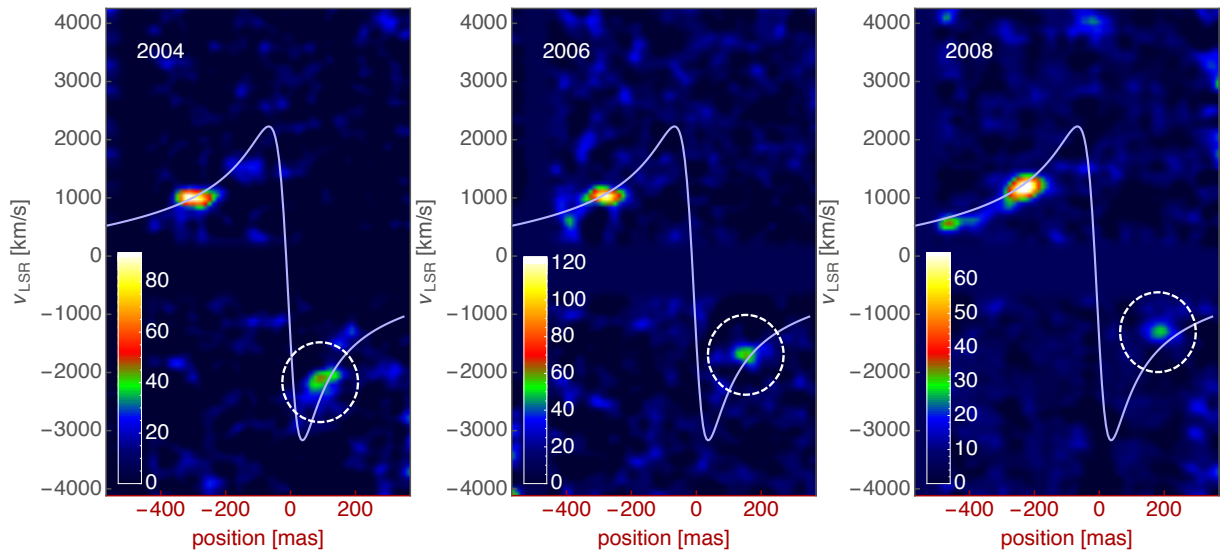

Figure 5: The pv-diagrams from the years 2004, 2006 and 2008 show a weak precursor, moving apparently along the orbit of G2. The scaling is adjusted in each observed map individually to optimally show the structure of the gaseous emission; the maps cannot be compared photometrically. The colorbar definition is the same as in Fig.1.

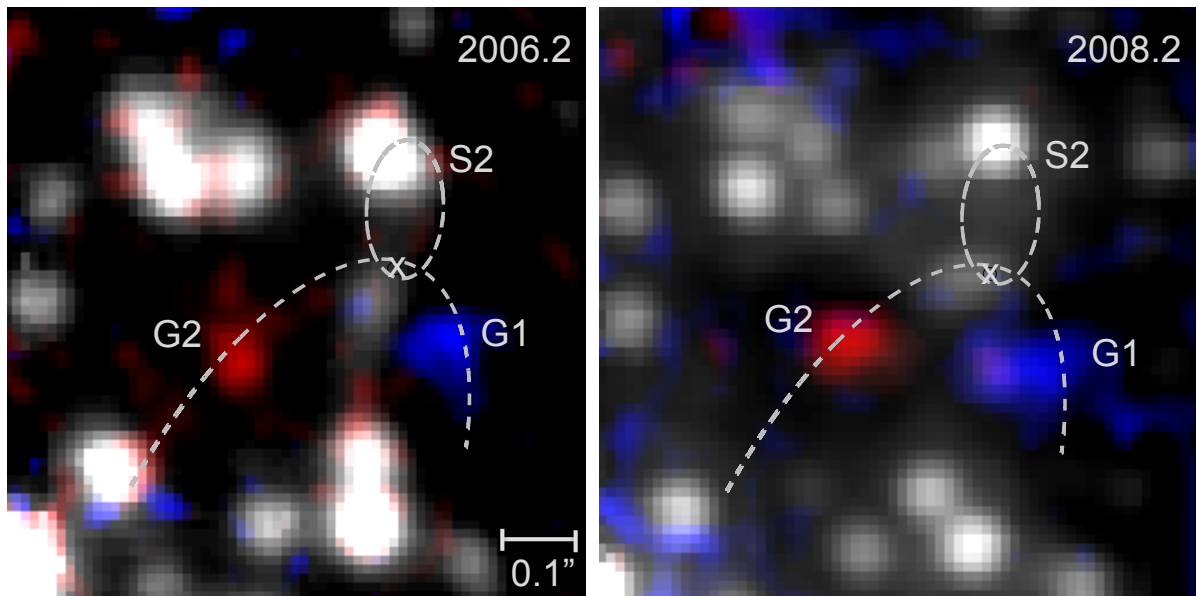

Figure 6: Color composites from our 2006 and 2008 data sets. The red channel is centered on the velocity of the Brackett-γ line peak of G2; the blue channel on the one of G1. The gray-scale background image is the continuum emission, showing the S-stars. The orbits of G2 and the star S2 are shown as dashed lines, and the position of Sgr A* is marked with a cross.

Although with large uncertainties of around 8 mas, we were able to determine the on-sky positions of this precursor from the Brackett-γ emission. To define the astrometric coordinates we use the nearby S-stars as a local reference. For those stars we know the astrometric positions from our K-band monitoring with high accuracy (Gillessen et al. 2009). The astrometric positions of the Brackett-γ precursor turn out to be broadly consistent with the positions of the L'-band source G1, which was first reported in Clénet et al. (2004a, 2005) and Ghez et al. (2005).

We were able to derive additional astrometric positions of G1 for seven epochs from 2003 to 2010 from NACO L'-band imaging. We used the tool Starfinder (Diolaiti et al. 2000) on the L'-band maps to retrieve the pixel positions (deconvolved maps are shown in figure 7) and applied the same technique of tying the coordinates locally to the S-stars. We assign errors of 3 mas to these positions, except for 2003 and 2004, for which we use 5 mas and 4 mas respectively to account for the potential confusion with Sgr A*.

Figure 7 also shows an M-band image from 2004 (Clénet et al. 2004b), in which G1's emission is also visible. On the other hand, we do not detect G1 in K-band to a limit of $m_K = 17.3$[2], nor are we aware of any K-band detection of it. This confirms its nature as a dusty, ionized gas cloud (Ghez et al. 2005). The total dust mass of G1 was estimated in Ghez et al. (2005) to be $1.3 \times 10^{-12}\, M_\odot$, very similar to the dust mass estimate for G2 of $\approx 10^{-12}\, M_\odot$[3]. During 2002, G1 was confused with Sgr A*. In the appendix A we show that heating up the dust content of G1 to $\approx 1200$ K could explain the mid-IR excess emission of Sgr A* in 2002 (Genzel et al. 2003).

From the total of ten astrometric positions (figure 8), we find that G1 exhibits an acceleration toward Sgr A* with a significance of 4.5σ. Together with the three radial velocity data points (which show a decrease from around 2000 km/s to 1100 km/s between 2004 and 2008), we have enough information to determine the orbit of G1. This assumes that indeed the gas emission seen as a precursor in our pv-diagrams can be associated with the L'-band source G1. We note that there is a systematic offset between the Brackett-γ and the L'-band derived positions; however a similar offset is present also in the G2 data (Phifer et al. 2013). Due to the small number of positions and unlike for G2, we cannot afford to separate the two data sets, and use thus both simultaneously for the fit. The best fitting orbital elements are given in table 1, along with the Brackett-γ orbit for G2 from Gillessen et al. (2013b) and the orientation of the clockwise stellar disk from Bartko et al. (2009). The orbit fit is shown in figure 8.

|      | semi-major axis | eccentricity | pericenter time | inclination | pos. angle. asc. node | long. pericenter |
|------|-----------------|--------------|-----------------|-------------|------------------------|------------------|
| G2   | 1.05 ± 0.25 "   | 0.976 ± 0.007 | 2014.25 ± 0.06 | 118 ± 2 °   | 82 ± 4 °               | 97 ± 2 °         |
| G1   | 0.36 ± 0.16 "   | 0.860 ± 0.050 | 2001.57 ± 0.40 | 108 ± 2 °   | 69 ± 5 °               | 109 ± 8 °        |
| Disk |                 |              |                 | 129 ± 18 °  | 98 ± 18 °              |                  |

Table 1: Comparison of the orbital elements of G1 and G2. The values for G2 are identical to the ones for the Brackett-γ orbit in Gillessen et al. (2013b). The orientation of the clockwise disk (CD) as published by Bartko et al (2009) is shown for comparison.

---

[2] We use the brightness of the star S55 / S0-102 (Meyer et al. 2012) as conservative limit. That star is comparable to G1 in terms of distance to Sgr A* and confusion with other sources. It is detected routinely in K-band, when not confused.

[3] The actual number given in the appendix of Gillessen et al. (2012) was incorrectly given as $10^{-10}\, M_\odot$, where a factor 100 was overlooked in the derivation.

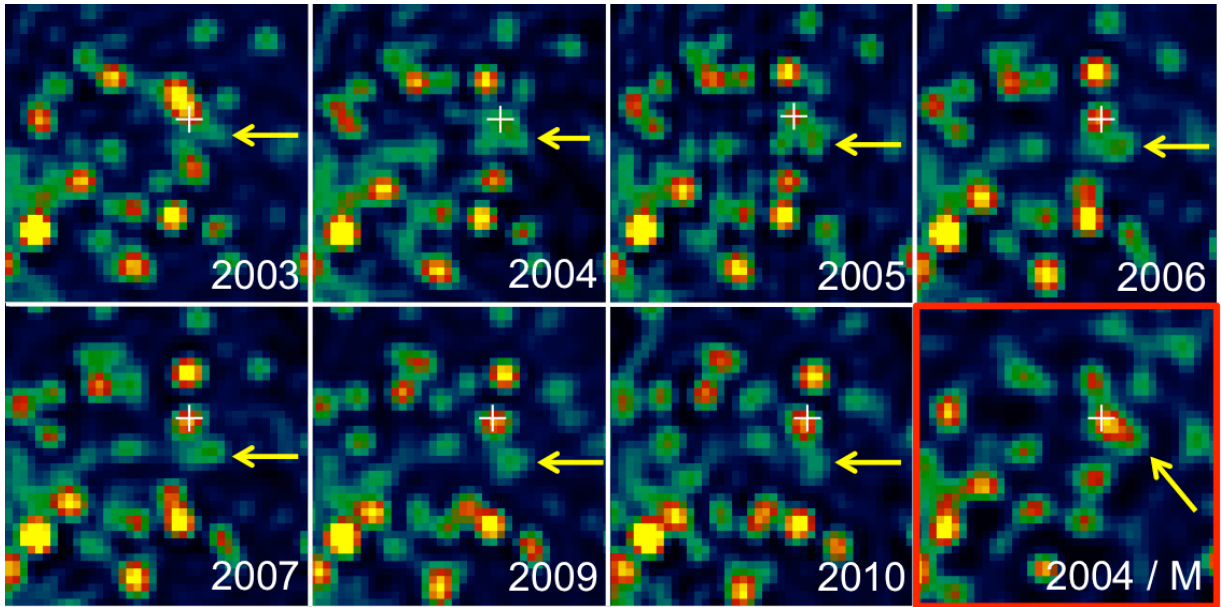

Figure 7: Series of deconvolved L'-band maps of the central arcsecond. The white cross marks Sgr A*. The yellow arrow points at the emission from G1. The bottom right panel is a deconvolved M-band map from 2004, showing also the G1 emission.

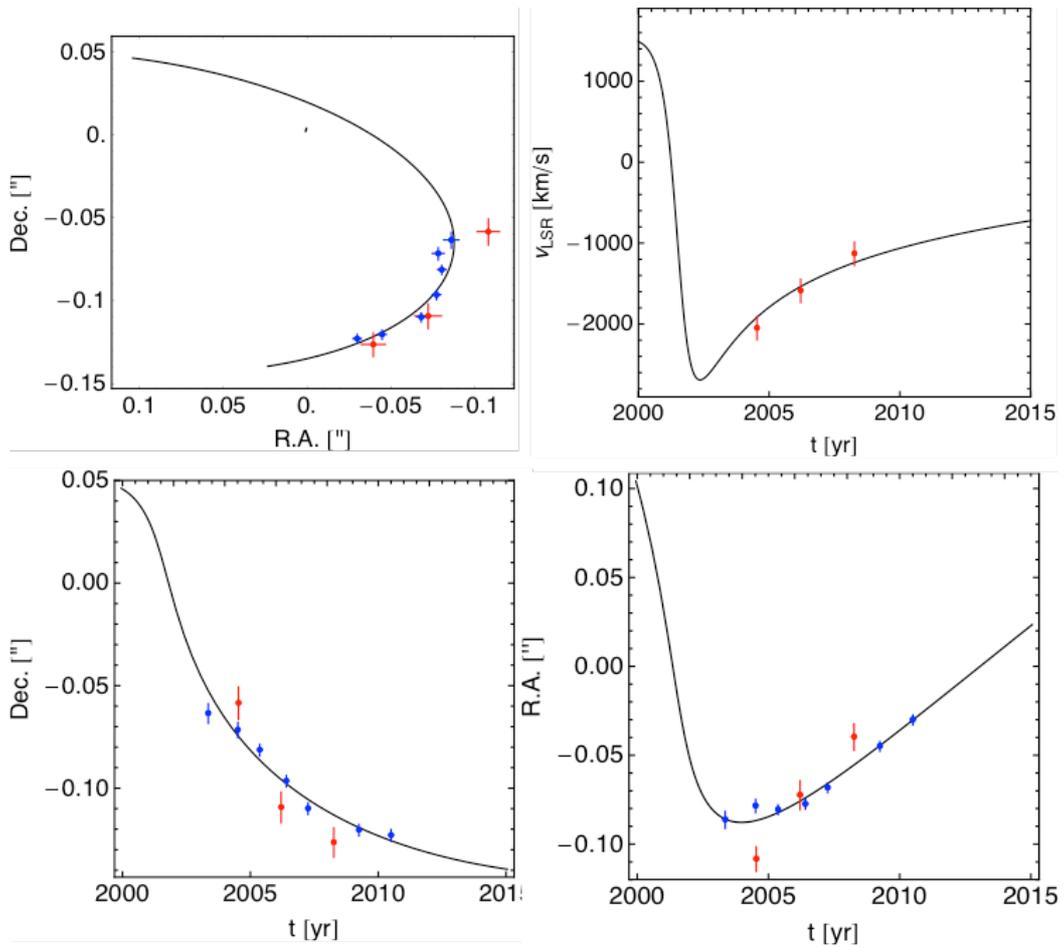

Figure 8: The orbit of G1. The blue data are L'-band astrometry derived from NACO. The red data are from SINFONI. The black line is the best fitting orbit.

The orientation of the G1 orbit is very similar to the G2 orbit in all three angles: G1 also lies in the plane of the clockwise stellar disk, and the orientation of the orbital ellipse in the plane is the same as for G2 (see table 1). The orbits differ in their semi-major axes and in their eccentricities. G1 is on a smaller orbit with lower eccentricity, and passed pericenter 12.8 years earlier. The fact that we see G1 after its pericenter passage is consistent with its diffuse appearance in the SINFONI data. The simulation work for G2 has shown that the passage of a gas cloud close to Sgr A* will result in a disrupted appearance a few years after pericenter (Anninos et al. 2012, Schartmann et al. 2012). Also the shape seen in the image presented by Ghez et al. (2005) supports that G1 has passed pericenter.

## *4. Discussion*

### *4.1 G2 luminosity*

The total brightness of G2 appears to have almost doubled between 2013 and 2014 from $2.3 \pm 0.6 \times 10^{-3}$ $L_\odot$ to $4.5 \pm 0.7 \times 10^{-3}$ $L_\odot$ (figure 2). A natural explanation for the increase might be the tidal compression, leading to a higher density of the gas in G2 and thus to an increase in line emission, which scales like density. The tidal compression acts in two dimensions, while the tidal shear only acts in one dimension, the combined effect of which is an increase in density during pericenter approach. The thermal pressure of the ambient medium might be important, too, as it increases inward. The flux increase might on the other hand also be a hint toward the models harboring a star in the center of G2, all of which predict a luminosity increase during pericenter (Murray-Clay & Loeb 2012, Scoville & Burkert 2013, Ballone et al. 2013). Yet, the increase predicted in Murray-Clay & Loeb (2012) is roughly a factor 5 over the time spanned by our data, more than what we observe.

### *4.2 Comparison of the tidal disruption with the test particle model*

In this section we show that the spatial-kinematic evolution of G2 continues to follow remarkably well what one would expect from the tidal disruption of an originally compact structure. In figure 9 we compare the time series of seven pv-diagrams with a test particle simulation. The gas cloud is modeled as an ensemble of non-interacting particles, each following its own Keplerian orbit around Sgr A*. The model shown is the same as in Gillessen et al. (2012), except for the orbital elements, which have been updated to the values from Gillessen et al. (2013b)[4]. The general growth of the tidal shear is captured. Also higher order features are reproduced, such as the small fraction of gas in the years 2012 and 2013 that overshoots the maximum redshift of the orbit. Similarly, the passage to the blueshifted side is described qualitatively nicely as well.

In the bottom right panel of figure 9, epoch 2014/04, one can also see a rim of emission connecting the two spots of the simulated emission at the redshifted and blueshifted side. Roughly 19% of the simulated particles are in that rim (defined by r < 50 mas and v < 2000 km/s). This emission is not detected in our data. This might simply be due to the low surface brightness, but it could also be a hint toward a clumpy structure of G2 (see also figure 3). On the other hand, the procedure by which we extract the pv-diagrams leads to a lower sensitivity to structure extended mainly along the velocity axis such as the rim in the simulation: We need to subtract from each column (and row) in the pv-diagrams the respective median. Row-wise (along the position axis) this corrects for the background varying for the different spectral channels. Column-wise (along the velocity axis) this removes artifacts from stellar, thermal emission that is present at any given position in each spectral channel. It is this step, which also decreases our sensitivity to vertical features, such as the rim would be.

---

[4] The cloud is simply set up as a spherical, Gaussian distribution of non-interacting particles that only feel the gravitational force of the black hole. The distribution has an initial velocity width of 120 km/s (FWHM) and a spatial extent of 42 mas (FWHM) per coordinate in the year 2000.0.

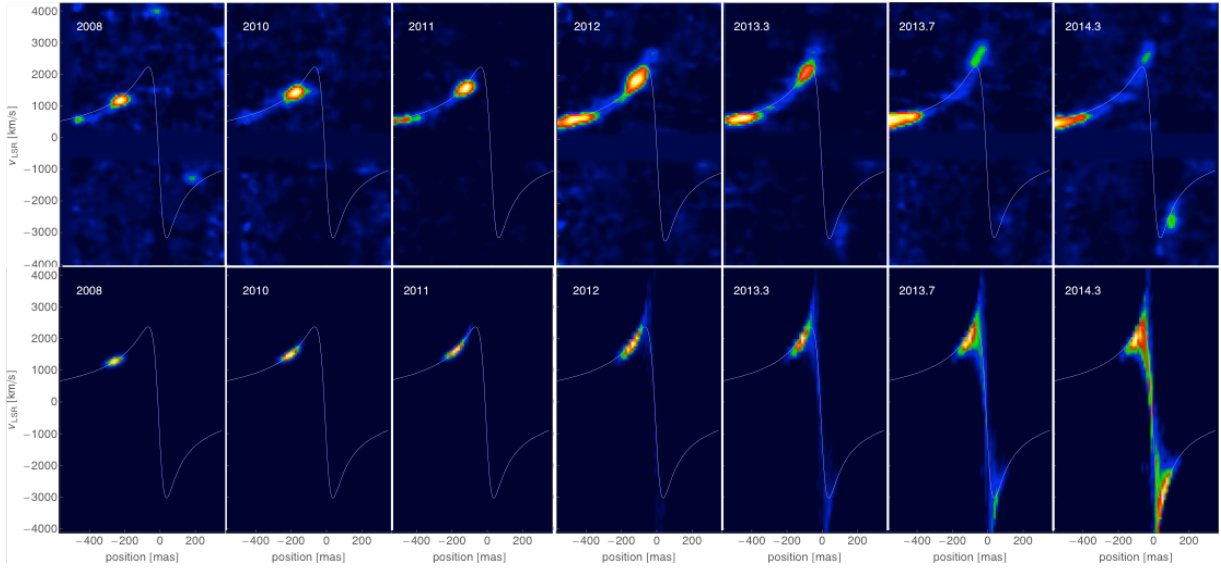

Figure 9: Comparison of the seven pv-diagrams from the epochs 2008, 2010, 2011, 2012, 2013/04, 2013/09 and 2014/04 (top row) with a test particle simulation (bottom row) of the same type as used in Gillessen et al. (2012). The scaling is adjusted in each observed map individually to optimally show the structure of the gaseous emission; the maps cannot be compared photometrically. The simulation plots show particle density.

Obviously, we could try to optimally match data and simulation by varying the initial conditions of our simulation. We did not attempt that yet for two reasons: a) instead of iterating the orbit of one particle during the fit, one would need to do that for thousands of particles; and b) the gain in insight is probably rather limited. The physics of the test particle model are necessarily incorrect, because it neglects external and internal hydrodynamic forces, which probably are relevant for G2 (Burkert et al. 2012). This type of quantitative comparison should therefore be done using hydrodynamic simulations, as in Schartmann et al. (2012). Also, we note that the evolution of the tail has not been at the focus of any of the simulations up to now.

When internal forces (i.e. self gravity or pressure) can be neglected, the tidal evolution on a highly elliptical orbit is dominated by a mode, which can be understood as two particles traveling on the same orbit with slightly different initial times. Following the arguments of Sari, Kobayashi & Rossi (2010) one can show that for an orbit seen from behind the length $l_p$ of the filament at pericenter will be

$$l_p = \frac{1}{5}\sqrt{16\,\delta r_0^2 + 9\,\delta v_0^2\, t_0^2}\sqrt{\frac{r_0}{r_p}},$$

where $\delta r_0$ is the initial size, $\delta v_0$ is the initial velocity dispersion, $r_0 \gg r_p$ is the starting point, $r_p$ is the pericenter distance, and $t_0 \sim r_0^{3/2}$ is the time to pericenter. The above formula is only correct for $l_p \ll r_p$, and will thus be only approximately correct for G2. Yet, the basic scalings remain the same also if $l_p > r_p$. Hence, there are potentially many combinations of $\delta r_0$, $\delta v_0$ and $r_0$ leading to similarly sized clouds at pericenter. This shows that there is not a unique set of initial conditions that can describe the data, and one could start the simulation also earlier than in 2000.0 and further out than currently assumed.

The simulations presented so far (Schartmann et al. 2012, Anninos et al. 2012, Ballone et al. 2013, Abarca et al. 2014, De Colle et al. 2014) justify the simple test particle model a posteriori: Consistently, the more involved calculations show that the evolution of G2 pre-pericenter is very similar to that of an ensemble of test particles. It is only at or after pericenter that the hydrodynamic effects become significant and eventually take over.

We can also use the test particle simulation to investigate the appearance of G2 in images (figure 10), allowing comparisons with imaging data. In order to compare with the observations reported in Ghez et al. (2014), we have taken our simulation at the epoch 2014/04, and projected the density into the celestial plane with a pixel size of 12.5 mas. After smoothing the result with a Gaussian kernel with 4 pixel radius (corresponding roughly to the reported resolution of 90 mas) we obtained an image (figure 10, right), which we fitted with an elliptical Gaussian. It has a size of 100 mas × 175 mas (FWHM). Thus, our simple model is only marginally compatible with the reported compactness of G2.

One way to improve the agreement, i.e. to get a more compact appearance at the pericenter epoch, might be to start the simulation earlier, with a smaller initial size instead. An earlier creation of G2 than 2000.0 seems more likely also for other reasons (Burkert et al. 2012); for example if it formed as a clump in a wind of one of the disk stars. For the given mass, a smaller cloud formed further out cannot be in pressure equilibrium, however.

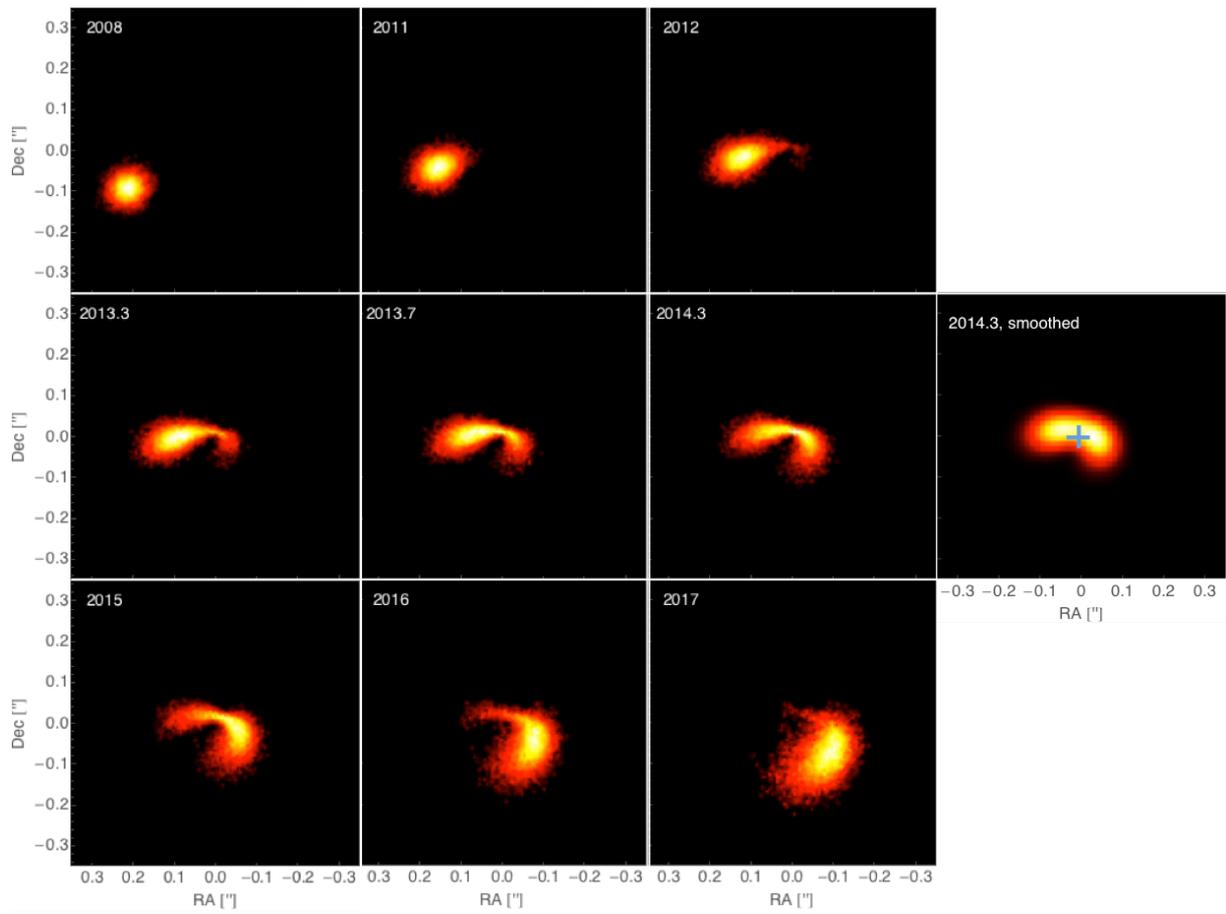

Figure 10: The same test particle simulation as in figure 9, projected into celestial coordinates relative to Sgr A*. The nine panels on the left use a square-root-scaling of the particle density (emphasizing thus fainter structures) and a smoothing kernel of 2 mas. The panel on the right shows in a linear scale how the 2014 epoch would look like for a resolution of 80 mas on a pixel grid with 12.5 mas/pix. The position of Sgr A* is marked with a cross.

*4.3 On the peri-center distance*

For an observer in the plane of the orbit, the overall change in radial velocity is $2GM_{BH}/L$, where $L$ is the angular momentum of the orbit[5,6]. For an observer out of the plane, only a component of this velocity change is observed $2GM \sin i /L$. For highly eccentric orbits, we can substitute $L = \sqrt{2GMr_p}$, where $r_p$ is the peri-center distance, and obtain

$$\frac{\Delta v}{c} = \sqrt{\frac{r_s}{r_p}} \sin i.$$

,where $r_s$ is the Schwarzschild radius. The inclination $\sin i$ is already known from the early portion of the orbit, where G2 had a well defined position. We can now use single images, where the cloud is widely spread and spans the full change in velocity $\Delta v$, to calculate $r_s/r_p$.

From Figure 1 we can extract the radial velocity change from a pre-peri velocity of 2700 +/− 150 km/s to a post-peri of −3300 +/− 150 km/s. If we take into account the best-fit inclination of $118^0$, then we can infer the peri-center distance of $r_p = 1950 +/−90\ r_s$. This value matches the predicted peri-center distance of $r_p = 2000\ r_s$ from Gillessen et al. 2013b.

Earlier estimates from Phifer et al. (2013) predicted a too close passage with only $1600\ r_s$.

This method presented here provides a neat tool to quickly assess the scales involved. In principle an orbital fitting also retrieves the peri-center distance. However, depending on how well the data samples the orbit, the eccentricity (i.e. peri-center distance) can be quite loosely constrained. Yet the inclination is often better constrained. In this case, the presented method yields a more accurate peri-center measurement, provided the velocity change is well captured.

---

[5] The simplest proof is to note that an elliptical orbit is a circle in the velocity plane: angular momentum is $L = r^2(d\theta/dt)$. Therefore, if we divide the ellipse into infinitesimal segments, each with opening angle $d\theta$, as viewed from the black hole, then each segment takes a time of $dt = r^2 d\theta/L$. Within this time the black hole changes the velocity by a constant amount $dv = GM\ dt/r^2 = GM\ d\theta/L$. The direction of the velocity change is along $\hat{r}$, therefore also changing its direction by $d\theta$ in each segment, and together portraying a circle of radius $v = \frac{dv}{d\theta} = GM/L$. The projection of this circle on the line of sight, which is the diameter of this circle $\Delta v = 2GM/L$, is the change in velocity the observer will see.

[6] A Second more direct proof is to note that the maximal and minimal velocities are obtained when the star-black hole-observer angle is a right angle. Therefore the change in the angle of the star as viewed from the black hole is $\pi$. If we take the minimal velocity to be at angle $\theta = 0$, and integrate the change in radial velocity $\int GMr^{-2} \sin \theta\ dt = \int_0^\pi GM/L\ \sin \theta\ d\theta = 2GM/L$.

*4.4 The tail emission*

We detect a tail emission in Brackett-γ and L'-band continuum, which seems to follow G2. The tail grows with time as G2 approaches the black hole. As already noted for the Brackett-γ emission in Phifer et al. (2013) and Gillessen et al. (2013a) we again find that the tail seems to be slightly offset from the G2 orbit both in the L'-band images and in the Brackett-γ maps. Possible reasons for that are (i) the uncertainty in our knowledge of the orbit of G2; (ii) as illustrated in Guillochon et al. (2014), a longer gas streamer does not necessarily line up along a single orbit, but can create a whole family of similar orbits; (iii) the tail might be affected by other forces than gravity; in particular one could imagine an outflow originating from Sgr A* dragging the tail away from the orbit, or friction with the ambient medium; and (iv) the tail is not associated to G2 and is a chance projection (Phifer et al. 2013). However three arguments render a chance association unlikely. First, the tail follows G2 in time. Second, it is continuously connected to the head, G2. And third, it shows a velocity gradient, consistent with the orbit. From this, it seems that the cloud G2 is actually a bright knot of a longer gas streamer.

## 5. A speculation: A connection between G2 and G1

The similarity of the G1 and G2 orbits in the three angles is astonishing, and we explore in the following whether there might be a physical connection between G1 and G2. The basic idea is that G1 and G2 might be clumps of the same gas streamer. In this case, both clouds would share a common orbit before pericenter, namely the one of G2. After pericenter passage however, the orbit is altered due to the interaction with the ambient medium. Is it possible that G1 has obtained a smaller semi-major axis and a lower eccentricity as a result of the pericenter passage in 2001?

*5.1 The ram pressure toy model*

To answer this question, we make the simple assumption that G1 was decelerated during pericenter passage by a drag force due to the ram pressure of the surrounding medium, and as a result was pushed onto a lower energy, more circular orbit. We implement such a force in our orbital fitting. The force is simply proportional to the ram pressure ≈ ρ(r) v$^2$ and it acts against the direction of instantaneous motion, not changing the orbital plane. The normalization of the force is the only additional free parameter in the fit. A constant normalization corresponds to a constant cross section, i.e. a constant cloud radius. We assume that the density of the ambient medium ρ(r) scales like r$^{-1}$ (Yuan et al. 2003, Xu et al. 2006). For a highly eccentric orbit and a density profile increasing inward, the effect of the drag force can also be understood as a quasi-instantaneous loss of energy and angular momentum at pericenter passage, and free-falling motion otherwise.

If the idea holds, we should be able to simultaneously fit the G2 and G1 data, when we take into account the different pericenter times. Thus, we add to the observation times of the G1 data the difference between the two pericenter times, i.e. 12.8 years, and obtain a combined data set that extends into the future. In this sense, G1 is assumed to give a preview of what will happen to G2.

The resulting fit is well constrained, although the reduced χ² has a value of 4.7 (figure 11). In particular, the fit misses the radial velocities of G1 somewhat, but it reproduces the general trend (i.e. the acceleration) and magnitude of these velocities. Given the simplicity of this approach, the agreement is nevertheless quite remarkable. The drag force parameter is dimensionless and the fit yielded a numerical value of (8.5 ± 1.4) × 10⁻³ (see section 5.2).

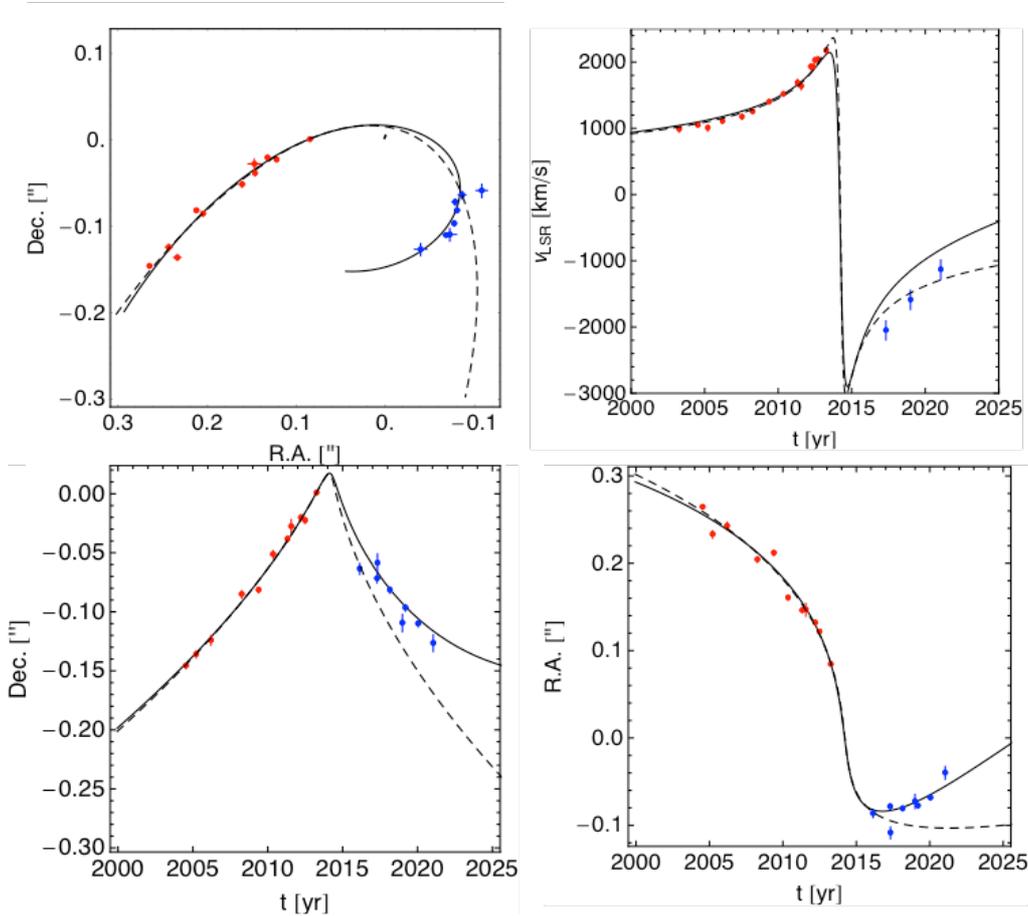

Figure 11: The combined orbital fit of the G2 and G1 data, using a drag force beside Newtonian gravity. The red data are the G2 data; the blue data are the G1 data, moved in time forward by 12.8 years. The black solid line indicates the best fitting orbit. The dashed line is the original G2 orbit. In the top left panel the effect of the drag force is apparent.

We have also implemented the power-law index of the density profile as a free parameter, such that we can solve for it with the fit. This would constitute a new route to constraining the density profile of the hot gas around Sgr A*. It turns out, however, that the parameter is not well constrained, and we thus keep our initial choice. Also, we checked whether we could improve the fit by allowing the time delay between the two data sets to vary. It turns out that the improvement would not be significant, and the time delay would change by only −0.4 years. Hence, again we stay with our original choice.

The orbit with the drag force can be approximated by a Keplerian approach, a short period of energy loss during pericenter, and Keplerian motion afterwards again. To check this asymptotic behavior and estimate the energy loss, we use the combined orbit (including the drag force) to create a simulated data set, of which we fit pre- and post-pericenter parts separately with purely Keplerian orbits. The two fits are presented in table 2. For the pre-pericenter part, we essentially retrieve back the original G2 orbit.

For the post-pericenter part, one gets an orbit similar to the G1 one, with the orbital angles agreeing, but with a smaller semi-major axis. This moderate mismatch is of course only another sign of G1's radial velocities not being well reproduced by the combined fit.

|  | semi-major axis | eccentricity | pericenter time | inclination | pos. angle. asc. node | long. pericenter |
|---|---|---|---|---|---|---|
| Pre-peri | 0.99" | 0.969 | 2014.35 | 116° | 86° | 96° |
| Post-peri | 0.20" | 0.851 | 2001.38* | 116° | 86° | 96° |

Table 2: Simulating data according to the combined G1-G2 orbit, and fitting the pre- and post-pericenter parts separately yields the two orbits given here. For such a simulated data set, it is meaningless to derive errors of the parameters, because they could be made arbitrarily small by simulating arbitrarily good data. (* The pericenter time is corrected for the offset of 12.8 years.)

We checked, whether some simple changes would resolve the imperfectly matching radial velocities of G1, but found no significant improvement:
- Not allowing for a drag force in the combined fit leads to an orbit that does not match the on-sky curvature of the G1 data (figure 11). The radial velocities for this fit are more negative and thus match the G1 data better, however, the rate of change of the radial velocities (i.e. the radial acceleration) is not matched.
- The shape of the pv-diagrams and the residuals of the G2 radial velocity data points suggest that an orbit with higher eccentricity and/or more edge-on inclination would allow for a better fit of the radial velocities of both G1 and G2. Also, the orbit given in Phifer et al. (2013) has a larger eccentricity. We tried to use such orbits by down-weighting the astrometric data compared to the radial velocity data. It turns out, that indeed the eccentricity is increased then. However, the combined fit still does not improve significantly.
- Similarly, we investigated whether the 2013 radial velocity creates a bias. One might suspect this, because some of the gas was already observed on the post-pericenter side at that epoch, causing a bias in the mean radial velocity of the redshifted gas. Also with that change, we were not able to improve the fit.
- We used the flatter density profile $\rho(r) \sim r^{-0.5}$ from Wang et al. (2013). Again, we find that the fit does not change significantly.

Another way to improve the fit might come from the observation that the gas of the accretion flow will not be at rest (as we have assumed), but is rather radially moving. This adds another, radial force component. We have implemented such a force, yielding another free fit parameter corresponding to the strength of the force. It is physically related to the cross section of the cloud and the velocity of gas in the accretion flow. Using that model, we were not able to improve the fit significantly. Also the additional parameter is only marginally determined, and we have to conclude that we are not sensitive to the effect. Obviously, if one would allow for a rotating accretion flow, one could add another three free fit parameters, because then the direction and strength of the additional force are free. We have not followed that route, since already one additional parameter was not well determined anymore by the data available.

*5.2. Interpretation of the toy model*

Our drag force model, which assumes standard ram pressure and a constant cloud cross section, captures the G1 and G2 orbits remarkably well. The agreement of the model with the data renders the idea that G1 and G2 are part of the same gas streamer highly plausible. This is illustrated in another way in figure 12, where we show a combination of the nine pv-diagrams. In this representation the emission of both G2 and G1 apparently trace a very similar orbit. Deviations from the fit as found in section 5.1 are not surprising considering the simplicity of the model, which is neglecting very likely essential physics.

The remaining fit residuals might simply be due to the fact that the description of the motions of gas clouds as point particles is not accurate enough. Also, it is possible that the G1 orbit was already different from the one of G2 before pericenter, yet the two clouds might belong to the same gas streamer. The situation might be similar to the connection between tail and G2, which apparently form a physically connected structure, yet the orbits appear to be slightly different.

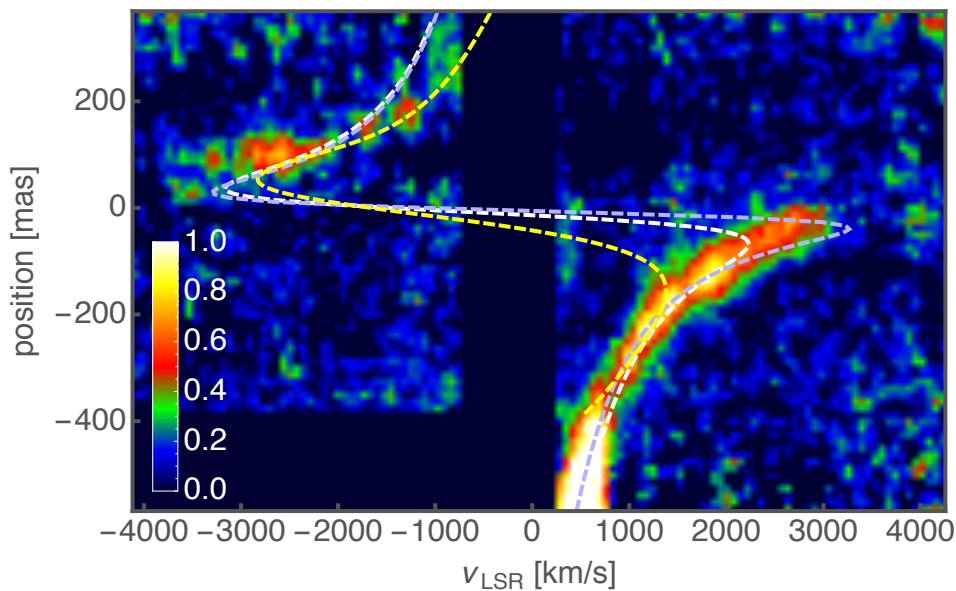

Figure 12: Position-velocity-diagram obtained by taking in each pixel the maximum of the ten values from the individual diagrams. On the blue shifted side, the diagrams containing flux (2004, 2006, 2008, 2013/04, 2013/09, 2014/04, 2014/07) have been given slightly higher weights than the other three diagrams. The intensity is square-root scaled. Since this does not represent a physical scale anymore, we only give relative, arbitrary intensity units. The white line is the Brackett-γ orbit from Gillessen et al. (2013b), the blue line the one from Phifer et al. (2013). The yellow line corresponds to the G1 orbit.

We have also checked, what the effect of the drag force of the given strength would be on the test particle simulation. Figure 13 shows the resulting pv-diagrams. Qualitatively, there is no difference visible to the Keplerian case shown in figure 9. The only noticeable difference is that the maxima of the radial velocities around pericenter passage are a bit less extreme, but the differences are significantly smaller than the intrinsic velocity spread, and also smaller than the differences between simulation and observation. The effect of the uncertainty in eccentricity exceeds thus the effect of the drag force. The situation changes when considering later points in time. In Appendix B we show the on-sky projection of the particle density for epochs up to the year 2030. One can see how an inspiral develops, but only after 2016. Clearly, this is a very uncertain approximation, since the hydrodynamic effects probably will get important after pericenter passage. The model would remain valid only if G2 had a very low volume-filling factor.

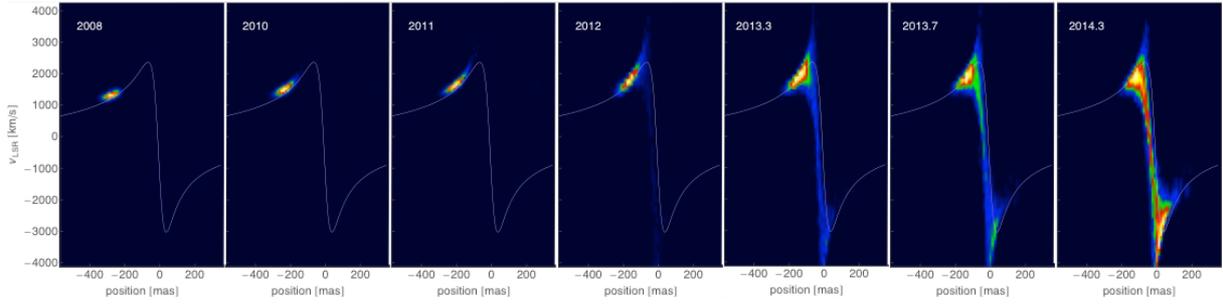

Figure 13: The same series of simulated pv-diagrams as in figure 9, however the test particles have been subject to the drag force as obtained by the combined G2-G1-fit.

The shape of the drag-force-orbit is at odds with what the hydrodynamic simulations have shown so far: Namely that G2 would rush through pericenter without much deflection. That behavior is not only due to the original density contrast between G2 and the hot gas, but also the result of the tidal compression, lowering the cross section of G2 as it approaches pericenter. We can ask, what the size of the cloud would need to be, for the given density profile of the hot atmosphere around Sgr A* to match the strength of the drag force as empirically determined from the combined fit.

The fit used a massless particle, such that force is per unit mass. The total force thus is

$$F_{\text{drag}} = a \frac{v^2}{r} M_{\text{G2}},$$

with $M_{\text{G2}} = 3\,M_{\text{Earth}}$ (Gillessen et al. 2012). The force is $\sim v^2$ and $\sim \rho(r) \sim r^{-1}$. The parameter $a$ determines the strength of the force and is dimensionless. The fit yielded $a = (8.5 \pm 1.4) \times 10^{-3}$. Approximating the cloud by a sphere of size s, the drag force will be (Murray & Lin 2004)

$$F_{\text{drag}} = \frac{\pi}{2} C\, s^2\, \rho(r)\, v^2.$$

For a compressible gas cloud one has $C \approx 1$. The ambient density profile we used is (Xu et al. 2006)

$$\rho(r) = 1.7 \times 10^{-20} \frac{\text{g}}{\text{cm}^3} \left(\frac{r}{10^{15}\text{cm}}\right)^{-1}.$$

Equating the two forces one can solve for *s* and finds $s \approx 2.4 \times 10^{15}$ cm. That corresponds very well to the size estimate for G2 by Gillessen et al. (2012), who measured an intrinsic size of $\approx 15$ mas = $1.8 \times 10^{15}$ cm for an assumed distance of $R_0 = 8.3$ kpc (Gillessen et al. 2009).

Hence, if G1 maintained roughly its original cross section during pericenter approach, the measured strength of the drag force matches what we would expect for the assumed density profile. The question thus is, whether the current simulations showing a decrease in cross section might not miss a key ingredient. Shcherbakov (2014) pointed out that G2 might be a magnetically arrested gas cloud. Recent work by McCourt et al. (2014) shows that an internal magnetic field of a gas cloud suppresses its disruption, while an external magnetic field increases the drag force as the cloud encounters and sweeps up magnetic field lines. One might thus speculate whether magneto-hydrodynamic simulations of G2 would yield a cross section larger than predicted by the current simulations, if not even constant, what in turn would lead to dragging effects similar to what we derived from the combination of G2 and G1 data.

## 6. Conclusion

We have presented observational evidence that the gas cloud G2 is actually part of a much larger gas streamer currently passing through the central arcsecond. There is gas in a tail of ionized gas and hot dust following G2 on an apparently very similar orbit. The tail is clearly connected with G2 in position and velocity, which makes previous claims of a chance association very unlikely. There is also evidence that about 13 years prior to the current pericenter approach of G2 another comparable gas/dust cloud, G1, went through pericenter on a similar orbit. A simple drag force model can decelerate a cloud on a G2-like orbit to an orbit that matches the G1 data. In this picture, G2 is thus a dense knot in a gas streamer, standing out by its compactness and brightness, and G1 is a precursing knot.

The idea of a gas streamer is at first glance compatible with the model in Guillochon et al. (2014), who propose that G2 is the tidal debris of a giant that underwent a partial disruption during its last pericenter passage at Sgr A*. Yet, this picture is not fully satisfying, since it does not explain naturally, why the G2 and G1 orbits are coplanar with the clockwise disk consisting of young O- and WR-stars. Instead, the idea of a clump in the wind of one of the massive disk stars appears more likely now again, since the main criticism to it was the compactness of G2. This problem is significantly weakened if G2 indeed is a gas streamer. It could then have formed around 100 years ago close to the apocenter of the current G2 orbit, without the unnatural need of a high initial velocity. The most likely stars for that scenario are S91 and IRS16SW. Both are found Southeast of Sgr A* at a projected distance of around 1'', and both have orbital phases which could be in accordance with the creation of G2 in the disk.

Another idea in the gas streamer picture is that it might explain the lack of evidence of strongly enhanced emission while G2 was plowing toward Sgr A*. Gillessen et al. (2012) and Sadowski et al. (2013a,b) estimated that the formation of a shock front might lead to substantial heating and observable X-ray emission or to particle acceleration and radio emission. The non-detection of such emission is not yet understood. In the gas streamer scenario, one can speculate whether the whole path of G2 is already filled with gas in pressure equilibrium with the ambient gas. This would suppress the formation of a shock. We also note that it is unlikely that G1 itself cleared the path for G2, since the time scale to refill the channel can be estimated from the size of G2 and the local speed of sound to be shorter than a year. This is much less than the temporal distance of G1 and G2 on the orbit.

### Acknowledgements
A.-M. M. is supported by the National Aeronautics and Space Administration through Einstein Postdoctoral Fellowship Award Number PF2-130095 issued by the Chandra X-ray Observatory Center, which is operated by the Smithsonian Astrophysical Observatory for and on behalf of the National Aeronautics Space Administration under contract NAS8-03060.

*Appendix A: The IR-excess of S2/Sgr A\* in 2002*

In 2002, the star S2 passed the pericenter of its orbit, where it was confused with Sgr A*. The combined brightness of S2/Sgr A* exceeded significantly the flux of S2 (Genzel et al. 2003). This flux increase in both K- and especially L'-band was seen in all 2002 images, and thus it is unlikely that it was due to the occasional flares of Sgr A* (Dodds-Eden et al. 2011). The K-band magnitude was brighter by roughly 0.3 mag (Gillessen et al. 2009) corresponding to a 32% increase or 4.5mJy for a dereddened flux density of 14mJy for S2 ($m_K$ = 14.1, $A_K$ = 2.42, $R_0$ = 8.3 kpc, Gillessen et al. 2009, Fritz et al. 2011). A few stars have been close in projection to S2 around 2002: S19 with $m_K$ = 16.0, S38 with $m_K$ = 17.0, and S54 with $m_K$ = 17.4. The total flux due to those stars is around 0.5 mJy only and cannot explain the observed excess.

Genzel et al. (2003) reported a total dereddened combined flux of 33mJy in L'-band for an assumed $A_L$ = 2.1. With the newer value $A_L$ = 1.09 from Fritz et al. (2011) that flux density is 13 mJy. The L'-band magnitude of S2 of 12.78 (Ghez et al. 2005) corresponds to a dereddened flux density of 5.2 mJy, such that the excess in L'-band is 7.8 mJy or about 1 magnitude.

The fact that G1 passed pericenter in 2001 might offer a new explanation for this excess. In the following we show that the excess can be explained by G1, if its dust was heated up to T = 1200K. Ghez et al. (2005) estimated the total mass of G1 from its (optically thin) dust emission to be $1.3 \times 10^{-10}$ $M_\odot$, assuming a gas to dust ratio of 100. One finds a very similar total mass of around $10^{-10}$ $M_\odot$ when asking the question, how many dust grains of an assumed size of 100 nm and with a temperature of 700K are needed in order to create the observed L'-band brightness of G1 of 1.2 mJy for an assumed gas to dust ratio of 100. Also the ansatz in the appendix of Genzel et al. (2003) leads to the conclusion that a total mass of around $10^{-10}$ $M_\odot$ and normal ISM properties lead to the observed flux from the embedded dust.

An optically thin, dusty cloud with a mass of $10^{-10}$ $M_\odot$ with a temperature of 1200K would create the observed L'-band excess of 7.8 mJy. In K-band that would lead to an excess of 4 mJy, well agreeing with the observations. What could lead to that heating?

We think that it is unlikely that a particular star would be responsible. The two stars which come closest to G1 during pericenter passage are the young, hot main sequence O-star S2 (Martins et al. 2008), and the giant S38. The latter reaches a minimum 3D distance to G1 of $7.7 \times 10^{15}$ cm, while S2 even reaches $5.7 \times 10^{15}$ cm. Both encounters occur around the epoch 2002.25. Given the luminosity of S2, it will thus completely dominate the heating. Assuming that the heating follows the law given by Scoville & Kwan (1976), a dust temperature of 1200K would only be reached for a distance of $\approx 10^{14}$ cm. This is much smaller than the closest approach. This argument obviously neglected the potentially large spatial extent of G1 during pericenter passage, but renders stellar heating implausible.

Hence, we think that in a scenario in which the excess emission in 2002 is due to G1, it is more likely that hydrodynamic effects lead to the additional energy input. An estimate of that is beyond the scope of this work.

## Appendix B: Test particle simulation with drag force

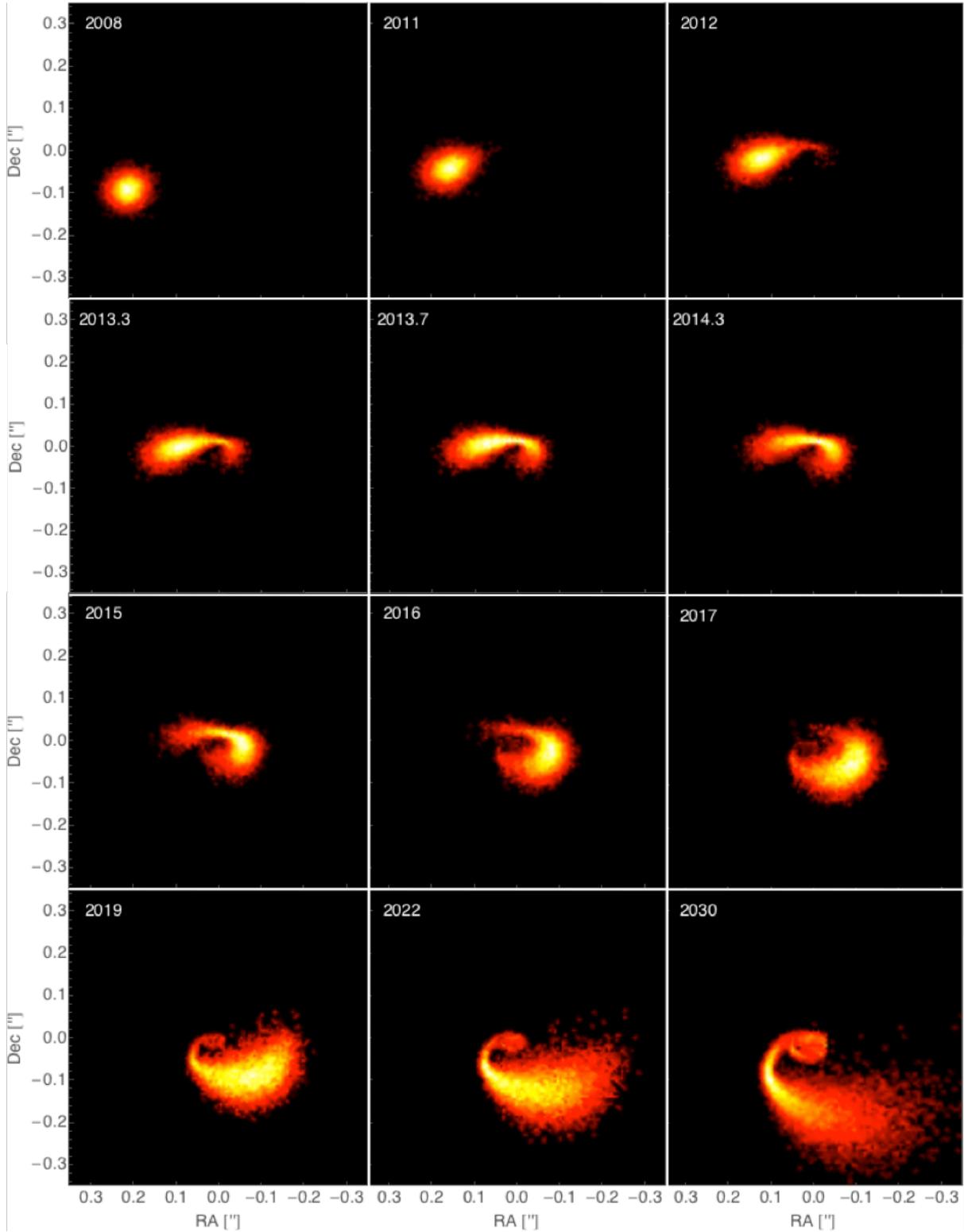

Figure 14: The on-sky projection of the test particle simulation in which the particles have been subject to the drag force as obtained by the combined G2-G1-fit. The color scale uses a square root scaling of the particle density.

## *Appendix C: Spectrum of G2 in April 2014*

The spectrum of G2 as seen in the deep integration in April 2014 is shown in Figure 15. The pre-pericenter redshifted Brackett-γ emission peaks at a velocity of 2800 km/s with a FWHM of 640 km/s. The post-pericenter emission peaks at a velocity of -2300 km/s with a FWHM of 850 km/s.

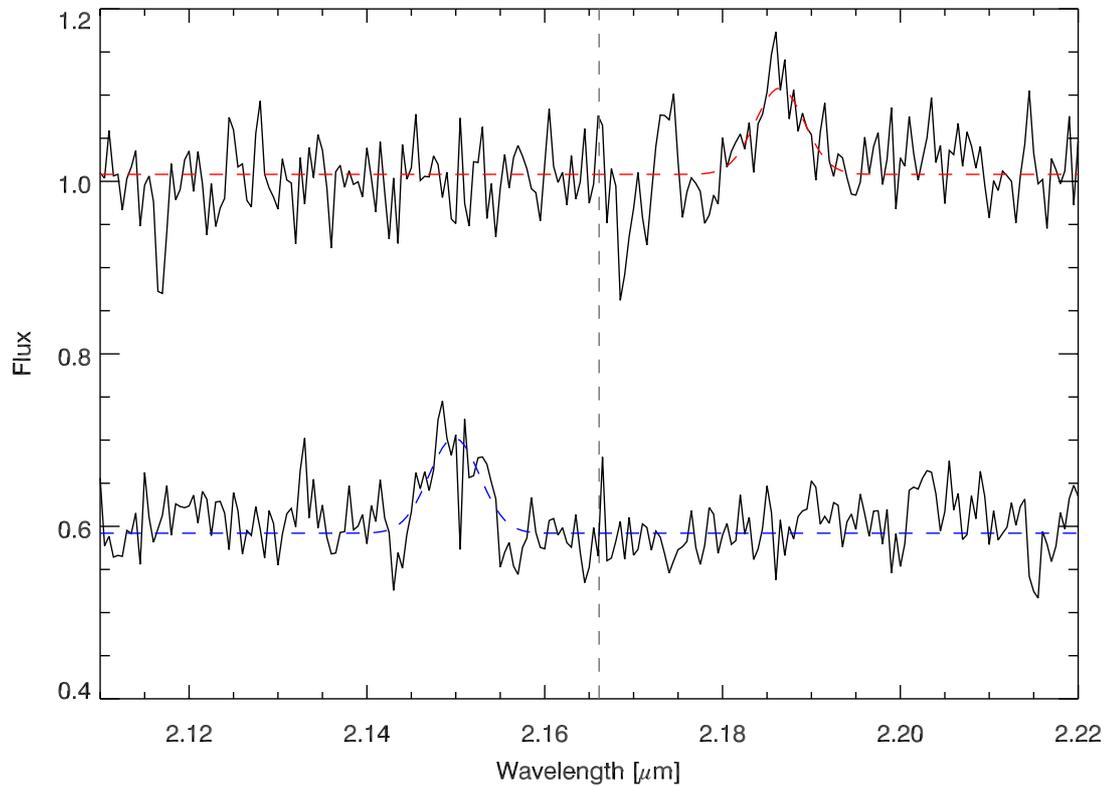

Figure 15: Spectrum of G2 pre-pericenter (top) and post-pericenter (bottom) as seen in April 2014. The vertical dashed line marks the rest-frame wavelength of Br-γ. Overplotted are Gaussian fits to the data.